\documentclass[12pt]{article}
\usepackage{amsmath}
\usepackage{graphicx}
\usepackage{subfigure}
\usepackage{hyperref}

\begin{document}

\author{Ernst Trojan and George V. Vlasov \and \textit{Moscow Institute of Physics
and Technology} \and \textit{PO Box 3, Moscow, 125080, Russia}}
\title{Acoustics of tachyon Fermi gas}
\maketitle

\begin{abstract}
We consider a Fermi gas of free tachyons as a continuous medium and find
whether it satisfies the causality condition. There is no stable tachyon
matter with the particle density below critical value $n_T$ and the Fermi
momentum $k_F<\sqrt{\frac 32}m$ that depends on the tachyon mass $m$. The
pressure $P$ and energy density $E$ cannot be arbitrary small, but the
situation $P>E$ is not forbidden. Existence of shock waves in tachyon gas is
also discussed. At low density $n_T<n<3.45n_T$ the tachyon matter remains
stable but no shock wave survives.
\end{abstract}


\section{Introduction}

Tachyons, first introduced for the description of superluminal motion \cite
{BDS62,F67}, are commonly known in the field theory as instabilities, whose
energy spectrum is 
\begin{equation}
\varepsilon _k=\sqrt{k^2-m^2}\qquad k>m  \label{tachy}
\end{equation}
where $m$ is the tachyon mass and relativistic units $c=\hbar =1$\ are used.
The concept of tachyon fields plays significant role in the modern research,
and tachyons are considered as candidates for the dark matter and dark
energy \cite{D1,D2}, they often appear in brane theories \cite{S1998} and
cosmological models \cite{FKS02,S2006}. A system of many tachyons can be
studied in the frames of statistical mechanics \cite{M84,DHR89}, and
thermodynamical functions of ideal tachyon Fermi and Bose gases are
calculated \cite{KRS07,KRS07b}.

In the present paper our interest is focused on the bulk and acoustic
properties of tachyon Fermi gas at zero temperature. Either it may concern
the Universe or a dense compact star, we consider the tachyon matter as
continuous medium and analyze its stability to the causality condition \cite
{EMM07,MV02}: 
\begin{equation}
c_s^2=\frac{dP}{dE}\leq 1  \label{caus}
\end{equation}
which implies that the sound perturbations must travel at a subluminal speed 
$c_s$. It is not evident whether the causality is satisfied at finite
density of tachyon matter.

The shock wave propagation in tachyon medium is another problem of our
interest. It is necessary to check the existence of stable shock waves and
find relevant characteristics of tachyonic medium. Applied problems of
astrophysics require more knowledge about collective features of tachyon
matter, and it is important to outline its non-trivial behavior.

\section{Tachyon Fermi gas}

Consider a system of free particles with the energy spectrum $\varepsilon _k$%
. The energy density of this system is defined as~\cite{Kapusta89}: 
\begin{equation}
E=\frac \gamma {2\pi ^2}\int \varepsilon _k\,f_k\,k^2dk  \label{erg}
\end{equation}
its pressure is

\begin{equation}
P=\frac \gamma {6\pi ^2}\int \frac{\partial \varepsilon _k}{\partial k}%
f_k\,k^3dk  \label{press}
\end{equation}
and the particle number density is 
\begin{equation}
n=\frac \gamma {2\pi ^2}\int f_k\,k^2dk  \label{dens}
\end{equation}
where $\gamma $ is the degeneracy factor. The distribution function of
fermions at zero temperature is approximated by the Heaviside step-function 
\begin{equation}
f_k=\Theta \left( \varepsilon _F-\varepsilon _k\right)  \label{hv}
\end{equation}
where $\varepsilon _F\equiv \varepsilon _{k=k_F}$ is the Fermi energy level
corresponding to the Fermi momentum $k_F$.

Substituting the energy spectrum of massive subluminal particles (bradyons) 
\begin{equation}
\varepsilon _k=\sqrt{k^2+m^2}  \label{free}
\end{equation}
in eq. (\ref{dens})-(\ref{press}) and integrating from $0$ to $k_F$, we find
the Fermi momentum 
\begin{equation}
k_F=\left( \frac{6\pi ^2n}\gamma \right) ^{1/3}  \label{f1}
\end{equation}
and the standard expressions for the energy density 
\begin{equation}
E=\frac \gamma {8\pi ^2}k_F^3\varepsilon _F+\frac 14mn_s  \label{erg1}
\end{equation}
and the pressure 
\begin{equation}
P=\frac \gamma {24\pi ^2}k_F^3\varepsilon _F-\frac 14mn_s  \label{press1}
\end{equation}
where 
\begin{equation}
n_s=\frac{\gamma m}{4\pi ^2}\left( k_F\varepsilon _F-m^2\ln \frac{%
k_F+\varepsilon _F}m\right)  \label{sc1}
\end{equation}
is the scalar density, and the Fermi energy is 
\begin{equation}
\varepsilon _F=\sqrt{k_F^2+m^2}  \label{fermi1}
\end{equation}

Substituting the energy spectrum of tachyons (\ref{tachy}) in eq. (\ref{dens}%
) and integrating from $m$ to $k_F$ we determine the Fermi momentum of
tachyons 
\begin{equation}
k_F=\left( \frac{6\pi ^2}\gamma n+m^3\right) ^{1/3}  \label{f2}
\end{equation}
Substituting (\ref{tachy}) in (\ref{erg}) and (\ref{press}), we obtain the
energy density of tachyons 
\begin{equation}
E=\frac \gamma {2\pi ^2}\int\limits_m^{k_F}\sqrt{k^2-m^2}k^2dk=\frac \gamma
{8\pi ^2}k_F^3\varepsilon _F-\frac \gamma {16\pi ^2}m^2\left( k_F\varepsilon
_F+m^2\ln \frac{k_F+\varepsilon _F}m\right)  \label{erg2}
\end{equation}
and the tachyon pressure 
\begin{equation}
P=\frac \gamma {6\pi ^2}\int\limits_m^{k_F}\frac{k^4dk}{\sqrt{k^2-m^2}}%
=\frac \gamma {24\pi ^2}k_F^3\varepsilon _F+\frac \gamma {16\pi ^2}m^2\left(
k_F\varepsilon _F+m^2\ln \frac{k_F+\varepsilon _F}m\right)  \label{press2}
\end{equation}
where the tachyon Fermi energy is 
\begin{equation}
\varepsilon _F=\sqrt{k_F^2-m^2}  \label{fermi2}
\end{equation}
In the light of (\ref{erg1})-(\ref{sc1}), we can define the tachyon scalar
density as 
\begin{equation}
n_s=\frac \gamma {4\pi ^2}m\left( k_F\varepsilon _F+m^2\ln \frac{%
k_F+\varepsilon _F}m\right)  \label{sc2}
\end{equation}
and rewrite (\ref{erg2})-(\ref{press2}) in a brief form like (\ref{erg1})-(%
\ref{press1}). This tachyon scalar density $n_s$ (\ref{sc2}) will be
incorporated in the energy density functional (\ref{erg2}) when one
considers a system of tachyons interacting via a scalar field. The
interaction is not so important in the ultrarelativistic tachyon gas and $%
n_s\rightarrow n$ when $k\gg k_F$.

At low density $n\rightarrow 0$ the formula (\ref{f2}) is expanded so 
\begin{equation}
k_F=m\left( 1+\frac{2\pi ^2}\gamma n\right)  \label{f22}
\end{equation}
When $k_F\rightarrow m$ the tachyon Fermi gas can be treated as
non-relativistic, and its Fermi energy (\ref{fermi2}) is approximated by
formula 
\begin{equation}
\varepsilon _F\cong 2\pi \sqrt{\frac n{\gamma m}}  \label{fermi3}
\end{equation}
Note that the relevant non-relativistic approximation of (\ref{fermi1}) for
the ordinary Fermi gas is written as 
\begin{equation}
\varepsilon _F=m+\frac 1{2m}\left( \frac{6\pi ^2n}\gamma \right) ^{2/3}
\label{fermi4}
\end{equation}
On the other hand, at high density 
\begin{equation}
n\gg \frac \gamma {6\pi ^2}m^3\qquad k_F\gg m  \label{hden}
\end{equation}
the formula (\ref{f2}) coincides with (\ref{f1}) and the formulas for
ordinary particles (\ref{erg1})-(\ref{press1}) and tachyons (\ref{erg2})-(%
\ref{press2}) yield the same ultrarelativistic EOS 
\begin{equation}
P=\frac E3  \label{ul}
\end{equation}

We can also present the formula (\ref{f2}) in the form 
\begin{equation}
n=\frac{\gamma m^3}{6\pi ^2}\left( \beta ^3-1\right)  \label{nb}
\end{equation}
with a dimensionless momentum 
\begin{equation}
\beta =\frac{k_F}m  \label{bet}
\end{equation}
Substituting it in (\ref{fermi2}), we write the Fermi energy in a universal
form 
\begin{equation}
\varepsilon _F=m\sqrt{\beta ^2-1}  \label{fermib}
\end{equation}
Substituting (\ref{bet}) and (\ref{fermib}) in (\ref{erg2}) and (\ref{press2}%
), we get universal formulas for the energy density and the pressure

\begin{equation}
E=\frac{\gamma m^4}{8\pi ^2}\left[ \beta ^3\sqrt{\beta ^2-1}-\frac 12\beta 
\sqrt{\beta ^2-1}-\frac 12\ln \left( \beta +\sqrt{\beta ^2-1}\right) \right]
\label{ergb}
\end{equation}
and 
\begin{equation}
P=\frac{\gamma m^4}{8\pi ^2}\left[ \frac 13\beta ^3\sqrt{\beta ^2-1}+\frac
12\beta \sqrt{\beta ^2-1}+\frac 12\ln \left( \beta +\sqrt{\beta ^2-1}\right)
\right]  \label{pressb}
\end{equation}

\section{Properties of stable tachyon matter}

The tachyon gas reveals most peculiar behavior at low density (\ref{nb}),
when $k_F\sim m$ ($\beta \sim 1$) and $P>E$. However, the stable continuous
medium must satisfy the causality condition (\ref{caus}). Otherwise, the
matter will be unstable to sound perturbations and may appear in the form of
droplets rather than continuous medium. For the ordinary Fermi gas (\ref
{erg1})-(\ref{press1}) it is satisfied automatically because 
\begin{equation}
c_s^2=\frac 13\frac{k_F^2}{k_F^2+m^2}\leq \frac 13\leq 1  \label{cs1}
\end{equation}
For the tachyon matter, in the light of (\ref{erg2})-(\ref{press2}), the
causality condition (\ref{caus}) implies 
\begin{equation}
c_s^2=\frac 13\frac{k_F^2}{k_F^2-m^2}\leq 1  \label{cs2}
\end{equation}
or, in a dimensionless form 
\begin{equation}
c_s^2=\frac 13\frac{\beta ^2}{\beta -1}\leq 1  \label{cs2b}
\end{equation}

Of course, each single tachyon moves faster than light, but existence of a
many-particle system of free tachyons does not contradict the causality (\ref
{caus}). The constraint (\ref{cs2}) implies that everywhere inside the
tachyonic medium it must be 
\begin{equation}
k_F\geq k_T=\sqrt{\frac 32}m  \label{ks}
\end{equation}
or, in terms of the dimensionless variable (\ref{bet}), it is 
\begin{equation}
\beta \geq \beta _T=\sqrt{\frac 32}\cong 1.225  \label{bets}
\end{equation}
Substituting the critical Fermi momentum $k_T$ (\ref{ks}) in (\ref{fermi2}),
we get the critical Fermi level 
\begin{equation}
\varepsilon _F\geq \varepsilon _T=\frac m{\sqrt{2}}  \label{fs}
\end{equation}
\textrm{\ }The Fermi energy of tachyon matter $\varepsilon _F$ cannot be
arbitrary small. The tachyon matter, satisfying the causality condition (\ref
{cs2}), is always relativistic matter, while the non-relativistic
approximation (\ref{fermi3}) could be applied only when $\varepsilon
_F\rightarrow 0$ ($\beta \rightarrow 1$).

Substituting the critical Fermi momentum (\ref{bets}) in (\ref{nb}),\ we
estimate the critical density 
\begin{equation}
n\geq n_T=\frac{\gamma m^3}{6\pi ^2}\left[ \left( \frac 32\right)
^{3/2}-1\right] \cong 1.41\times 10^{-2}\gamma m^3  \label{ns}
\end{equation}
The causality condition (\ref{cs2})\textrm{\ }implies that the tachyon
matter becomes unstable at low density $n<n_T$. The energy and pressure at
the causal point $\beta =\beta _T$ (\ref{bets}) are calculated according to
formulas (\ref{ergb})-(\ref{pressb}) so:

\begin{equation}
E_T=E\left( k_T\right) =\frac{\gamma m^4}{16\pi ^2}\left[ \sqrt{3}-\ln
\left( \sqrt{\frac 32}+\frac 1{\sqrt{2}}\right) \right] =6.80\times
10^{-3}\gamma m^4  \label{es}
\end{equation}
and 
\begin{equation}
P_T=P\left( k_T\right) =\frac{\gamma m^4}{16\pi ^2}\left[ \sqrt{3}+\ln
\left( \sqrt{\frac 32}+\frac 1{\sqrt{2}}\right) \right] =1.51\times
10^{-2}\gamma m^4  \label{ps}
\end{equation}
Hence, the tachyon pressure exceeds twice the energy density 
\begin{equation}
P_T\cong 2.23E_T  \label{point}
\end{equation}
It confirms the hypothesis of an EOS 
\begin{equation}
P>E  \label{hyp}
\end{equation}
that may not break the causality \cite{EMM07,MV02}. Equation 
\begin{equation}
E_1=E\left( k_1\right) =P\left( k_1\right)   \label{abs}
\end{equation}
determines 
\begin{equation}
\beta _1=\frac{k_1}m\cong 1.529  \label{k1}
\end{equation}
while 
\begin{equation}
\frac{E_1}{E_T}\cong 5.135  \label{sten}
\end{equation}
According to the equation of state (Fig. \ref{fig1}) the tachyon material
remains ''hyperstiff'' (\ref{hyp}) when the Fermi momentum varies in the
range of 
\begin{equation}
k_T<k_F<k_1  \label{range}
\end{equation}
In the light of (\ref{abs}) we may expect that the sound speed at the point
of ''stiffness'' $k=k_1$ should be $c_s=1$. However, substituting (\ref{k1})
in (\ref{cs2}) we find 
\begin{equation}
c_s^2\left( k_1\right) =0.58  \label{cs3}
\end{equation}
From eqs. (\ref{nb}), (\ref{ns}) and (\ref{k1}), we also find the particle
number density\ 
\begin{equation}
n_1=n\left( k_1\right) =0.0434\gamma m^3=3.04n_T  \label{n1}
\end{equation}

The equation of state of tachyon Fermi gas, calculated according to formulas
(\ref{ergb}) and (\ref{pressb}), is shown in Fig.~\ref{fig1}. The tachyon
matter behaves like ordinary ultrarelativistic Fermi gas (\ref{ul}) when its
parameters are much greater than the critical values (\ref{ns})-(\ref{ps}),
namely, when 
\begin{equation}
\bar n=\frac n{n_T}\gg 1\qquad \bar E=\frac E{E_T}\gg 1\qquad \bar P=\frac
P{P_T}\gg 1  \label{high}
\end{equation}
The ''stiffness'' of tachyon matter immediately increases at low density: it
becomes ''absolute stiff'' $E=P$ when $n=n_1=3.04n_T$ (\ref{n1}), and in the
range of densities $n_1<n<n_T$ it is even ''hyperstiff'' (\ref{hyp}). The
sound speed (\ref{cs2b}) vs the energy density (\ref{ergb}) is plotted in
Fig.~\ref{fig2}. It also reveals a rather peculiar behavior: it decreases at
high $E$ and tends to $1$ when the energy density approaches the critical
value $E\rightarrow E_T$ (\ref{es}).

\section{Shock waves in tachyon matter}

As soon as we have tested the causality condition in tachyon continuous
medium, it is reasonable to consider shock waves. The shock wave velocity $%
w_{-}$ and the velocity behind the shock $w_{+}$ are given by formulas \cite
{Taub78} 
\begin{equation}
v_{-}^2=\frac{P_{+}-P_{-}}{E_{+}-E_{-}}\frac{E_{+}+P_{-}}{E_{-}+P_{+}}
\label{sw1}
\end{equation}
and 
\begin{equation}
v_{+}^2=\frac{P_{+}-P_{-}}{E_{+}-E_{-}}\frac{E_{-}+P_{+}}{E_{+}+P_{-}}
\label{sw2}
\end{equation}
where index ''$-$'' indicates parameters before the shock wave, while index
''$+$'' corresponds to parameters behind it. Existence of stable shock waves
is determined by the constraint called as the evolutionary condition \cite
{LL87} 
\begin{equation}
v_{-}>c_{-}  \label{ev1}
\end{equation}
\begin{equation}
v_{+}<c_{+}  \label{ev2}
\end{equation}

For a shock wave of small amplitude 
\begin{equation}
v_{+}^2\rightarrow v_{-}^2\rightarrow v_{-}^2\rightarrow c_{-}^2=\frac{dP}{dE%
}  \label{sw0}
\end{equation}
we can apply a linear approximation and consider 
\begin{equation}
c_{+}^2=c_{-}^2+\frac{dc_{-}^2}{dE}\Delta E+...  \label{lin}
\end{equation}
where the increment 
\begin{equation}
\Delta E=E_{+}-E_{-}\rightarrow 0  \label{inc}
\end{equation}
is small. Eqs. (\ref{sw1}) and (\ref{sw2}), then, yield, respectively 
\begin{equation}
v_{-}^2\simeq c_{-}^2\left( 1+\frac{1-c_{-}^2}{E_{-}+P_{-}}\Delta E\right)
\label{sw11}
\end{equation}
and 
\begin{equation}
v_{+}^2\simeq c_{-}^2\left( 1-\frac{1-c_{-}^2}{E+P}\Delta E\right)
\label{sw22}
\end{equation}
Substituting (\ref{sw11}) in the evolutionary condition (\ref{ev1}), we have
to state that 
\begin{equation}
\Delta E>0  \label{pos}
\end{equation}
Substituting (\ref{sw11}) and (\ref{lin}) in (\ref{ev2}), we get 
\begin{equation}
1-\frac{1-c_{-}^2}{E_{-}+P_{-}}\Delta E<c_{-}^2+\frac{dc_{-}^2}{dE}\Delta E
\label{ev3}
\end{equation}
In the light of (\ref{pos}), the constraint (\ref{ev3}) implies that stable
shock waves propagate in the matter if 
\begin{equation}
\frac{E+P}{c_s^2\left( 1-c_s^2\right) }\left( -\frac{dc_s^2}{dE}\right) <1
\label{ev4}
\end{equation}
here we omit index ''$-$'' for simplicity and put $c_{-}\equiv c_s$.

The condition (\ref{ev4}) is automatically satisfied when 
\begin{equation}
\frac{dc_s^2}{dE}>0  \label{cp}
\end{equation}
that corresponds to ordinary matter. According to our calculation in Fig.~%
\ref{fig2}, the tachyon gas has always 
\begin{equation}
\frac{dc_s^2}{dE}<0  \label{cn}
\end{equation}
Substituting (\ref{cn}) in (\ref{ev4}), we have 
\begin{equation}
\frac{E+P}{c_s^2\left( 1-c_s^2\right) }\left| \frac{dc_s^2}{dE}\right| <1
\label{ev55a}
\end{equation}
In the light of (\ref{cs2}) and (\ref{erg2}) the inequality (\ref{ev55a}) is
developed so 
\begin{equation}
\frac{E+P}{c_s^2\left( 1-c_s^2\right) }\left| \frac{dc_s^2}{dk_F}\right|
\left( \frac{dE}{dk_F}\right) ^{-1}<1  \label{ev55}
\end{equation}
or 
\begin{equation}
\frac{E+P}{c_s^2\left( 1-c_s^2\right) }\left| \frac{dc_s^2}{d\beta }\right|
\left( \frac{dE}{d\beta }\right) ^{-1}<1  \label{ev55c}
\end{equation}
where the energy density $E$ (\ref{ergb}) and pressure $P$ (\ref{pressb})
are expressed in terms of dimensionless momentum $\beta =k_F/m$. Numerical
simulation shows that the constraint (\ref{ev55})-(\ref{ev55}) is satisfied
when 
\begin{equation}
k_F>k_D\cong 1.581m  \label{kd}
\end{equation}
\textrm{\ }The value $k_D$ exceeds the critical Fermi momentum $k_T=1.225$ (%
\ref{ks}). According to (\ref{nb}) and (\ref{ns}), the relevant particle
number density 
\begin{equation}
n_D=n\left( k_D\right) =3.45n_T  \label{nnd}
\end{equation}
sufficiently exceeds the critical density $n_T$. The relevant energy density
(\ref{ergb}) is 
\begin{equation}
E_D=E\left( k_D\right) =6.25E_T  \label{end}
\end{equation}
and the relevant sound speed (\ref{cs2}) is 
\begin{equation}
c_s^2\left( k_D\right) =0.56  \label{cs4}
\end{equation}
Formula (\ref{kd})-(\ref{end}) determine the band of shock wave instability
in tachyonic continuous medium 
\begin{equation}
k_D>k_F>k_T\qquad n_D>n>k_T\qquad E_D>E>E_T  \label{inst}
\end{equation}
In Fig.~\ref{fig2} it is labeled by shading. No discontinuity is possible
when the parameters of tachyon gas vary in the range (\ref{inst}). If,
anyhow, an abrupt peak of density is created, it will be transformed in a
smooth transition, rather than exist in the form of shock wave.

Substituting (\ref{pos}) and (\ref{cn}) in (\ref{lin}) we find that 
\begin{equation}
c_{+}<c_{-}  \label{string}
\end{equation}
This strange inequality (\ref{string}) is also applied to discontinuities of
the current in superconducting cosmic strings in the ''electric'' regime 
\cite{TV2011a}, while it is always 
\begin{equation}
c_{+}>c_{-}  \label{string2}
\end{equation}
for shock waves propagating in ordinary continuous medium which is
characterized by inequality (\ref{cp}).

\section{\textrm{\ }Conclusion}

\begin{figure}[tbp]
\caption{The ratio of pressure to energy density $P/E$\ vs Fermi momentum of
tachyon gas $\beta =k_F/m$}
\label{fig1}{\includegraphics[scale=0.4]{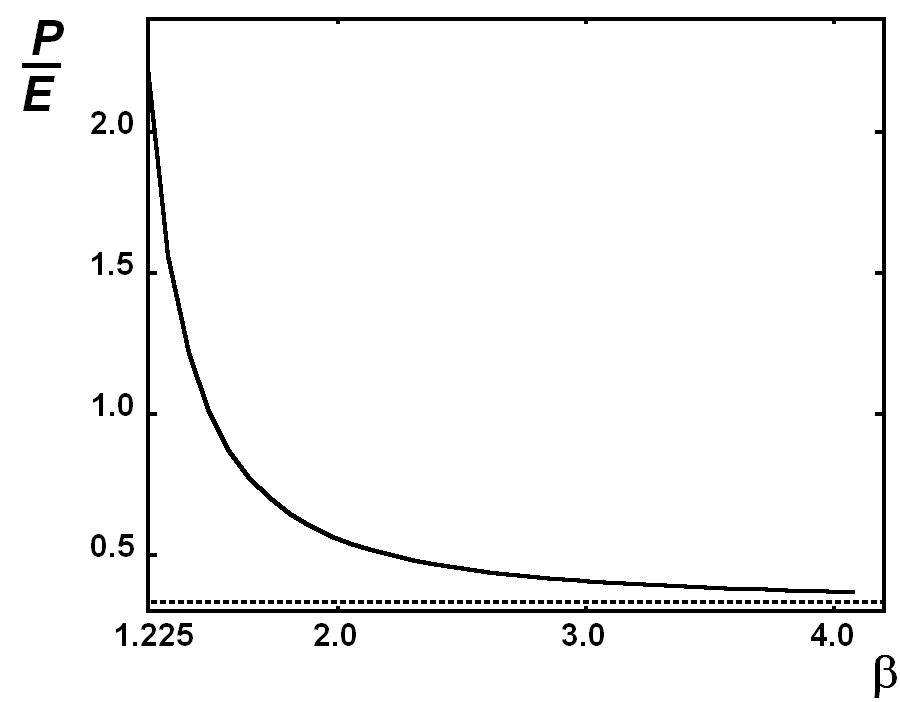}}
\par
Tachyon matter is unstable at $\beta <\sqrt{3/2}\cong 1.225$.
\par
Dotted line corresponds to ultrarelativistic equation of state $P/E=1/3$.
\end{figure}

\begin{figure}[tbp]
\caption{Sound speed $c^2_s$ vs energy density of tachyon Fermi gas $\bar
E=E/E_T$ [critical value $E_T$ is defined in eq.~(\ref{es})]}
\label{fig2}{\includegraphics[scale=0.4]{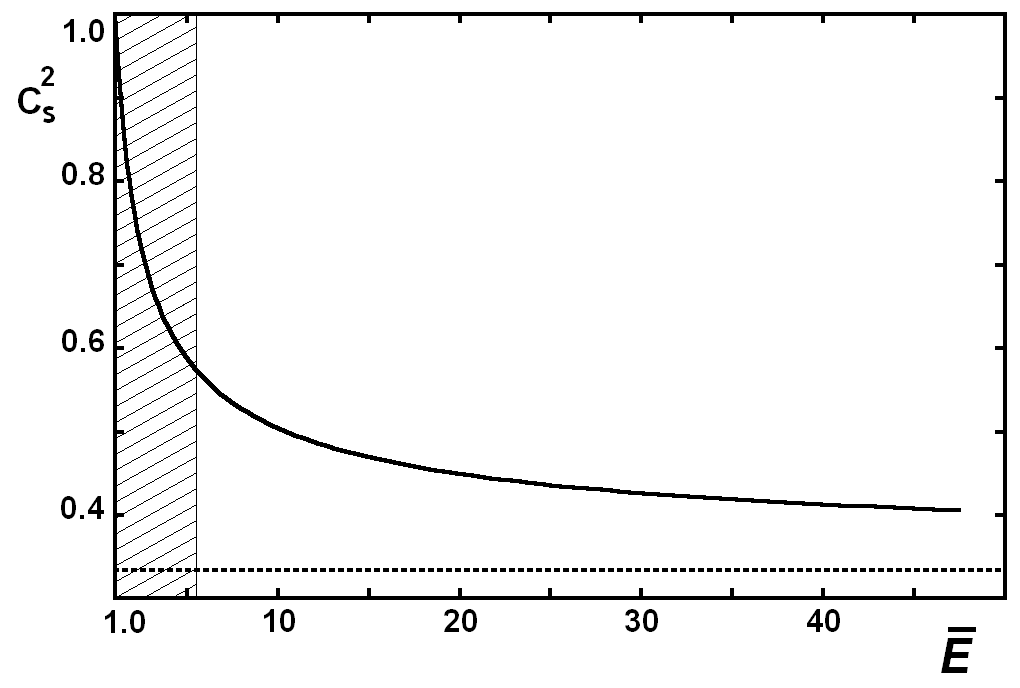}}
\par
Shading labels the domain where tachyon gas remains stable but no shock wave
is possible.
\par
Dotted line corresponds to the sound speed in ultrarelativistic matter $%
c^2_s=1/3$.
\end{figure}

A system of many tachyons with energy spectrum (\ref{tachy}) and the mass $m$%
, described in the frames of statistical thermodynamics and mechanics of
continuous medium, is endowed with unusual properties. The equation of state
(EOS) of tachyon Fermi gas at zero temperature is determined by eqs. (\ref
{erg2})-(\ref{press2}) or (\ref{ergb})-(\ref{pressb}) that is shown in Fig.~%
\ref{fig1}. The causality condition (\ref{caus}) is satisfied when the Fermi
momentum of tachyon gas is not less than the critical value $k_F\geq k_T=%
\sqrt{3/2}m$ (\ref{ks}). It implies that the tachyon continuous medium must
have finite density $n>n_T$ and no stable matter exists below the critical
density $n_T$ whose value depends only on $m$ (\ref{ns}). The peculiar
behavior of tachyon gas is seen especially near the critical point $%
k_F\rightarrow k_T$, particularly, its EOS becomes ''hyperstiff'' $P>E$
(see~Fig.~\ref{fig1}), while the sound speed $c_s$ decreases with the growth
of energy density (Fig. \ref{fig2}). If the tachyon material is still
expanded below the critical density $n_T$, it will lose stability to sound
perturbations and, perhaps, will appear in the form of dense droplets rather
than continuous substance.

Another peculiar property of tachyonic continuous medium concerns the shock
waves. The criterion of shock wave stability (\ref{ev1})-(\ref{ev2}) is
satisfied when the Fermi momentum exceeds the value $k_D=1.581m$ (\ref{kd}).
At the tachyon gas with $k_F>k_D$ is able to conduct shock waves. No no
shock wave can appear in the band of instability $k_D>k_F>k_T$. (\ref{inst}%
). If any discontinuity of density (pressure) is created artificially, it
will be unstable and decay.

The tachyon equation of state (Fig.~\ref{fig1}) is very ''soft'' ($P\simeq
E/3$) at high energy density $E\gg E_T$ (when $\beta =k_F/m\gg 1$) but it
becomes ''hyperstiff'' ($P>E$) as soon as the energy density approaches the
critical value $E_T$ (\ref{es}) when $\beta \rightarrow 1.225$. Therefore, a
star with tachyon content will have rather ''soft'' core and much more
''stiff'' envelope. Particularly, for the pi-meson mass $m=m_\pi =138\,%
\mathrm{MeV}$\ and at $\gamma =1$ the critical parameters of tachyon matter (%
\ref{ns}), (\ref{es}) are estimated as 
\begin{equation}
n_T=0.029n_0=0.005\,\mathrm{fm}^{-3}\qquad E_T=2\times 10^{-3}\rho _0
\label{nsp}
\end{equation}
where $\rho _0=2.8\times 10^{14}\,\mathrm{g\cdot cm}^{-3}=159~\mathrm{%
MeV\cdot fm}^{-3}$\ is the normal nuclear density and $n_0=0.166\,\mathrm{fm}%
^{-3}$ is the relevant particle density. For the nucleon mass $m=m_p=939\,%
\mathrm{MeV}$\ the critical parameters of tachyon matter are 
\begin{equation}
n_T=9.2n_0=1.53\,\mathrm{fm}^{-3}\qquad E_T=4.3\rho _0  \label{nsn}
\end{equation}
It brings more intrigue to the problem. The latter value of critical energy $%
E_T$ can be compared with the energy density $E_{*}$ in the center of a
neutron star with regular nuclear equation of state \cite{TV2010}, where it
is typically $E_{*}\sim 700~\mathrm{MeV\cdot fm}^{-3}\sim 5\rho _0$
(corresponds to the nucleon number density $n_{*}\sim 8n_0$).

It should be also noted that the star cannot contain tachyon matter if its
central energy density is smaller than $E_T$\ (\ref{es}). We may imagine a
tachyon core only at $E_{*}>E>E_T$\ because the tachyon gas cannot have free
surface with zero pressure $P=0$\ and its pressure must be always higher
than the critical pressure $P_T$\ (\ref{ps}). However, as soon as the energy
density $E>E_1=5.13E_T$ (\ref{sten}), the EOS\ of tachyon gas is even
''stiffer'' than the ''absolute stiff'' EOS $P=E$. Will the mass of tachyon
star be great? Is it possible to form supermassive stellar objects? It is
the subject for further research.

We are grateful to Erwin Schmidt for critical comments.

\newpage

\end{document}